**Terahertz Quantum Sensing**

Mirco Kutas[1,2]*†, Björn Haase[1,2]†, Patricia Bickert[1], Felix Riexinger[1,2], Daniel Molter[1], Georg von Freymann[1,2]*

[1]Fraunhofer Institute for Industrial Mathematics ITWM, Fraunhofer-Platz 1, 67663 Kaiserslautern, Germany.
[2]Department of Physics and Research Center OPTIMAS, Technische Universität Kaiserslautern (TUK), 67663 Kaiserslautern, Germany.
*Correspondence to: mirco.kutas@itwm.fraunhofer.de or georg.von.freymann@itwm.fraunhofer.de
† The first two authors contributed equally to this work.

**Abstract**

Quantum sensing is highly attractive for accessing spectral regions in which the detection of photons is technically challenging: sample information is gained in the spectral region of interest and transferred via biphoton correlations into another spectral range, for which highly sensitive detectors are available. This is especially beneficial for terahertz radiation, where no semiconductor detectors are available and coherent detection schemes or cryogenically cooled bolometers have to be employed. Here, we report on the first demonstration of quantum sensing in the terahertz frequency range in which the terahertz photons interact with a sample in free space and information about the sample thickness is obtained by the detection of visible photons. As a first demonstration, we show layer thickness measurements with terahertz photons based on biphoton interference. As non-destructive layer thickness measurements are of high industrial relevance, our experiments might be seen as a first step towards industrial quantum-sensing applications.

**Introduction**

In the last decade, quantum sensing and imaging has become a popular scheme for measurements in the infrared spectral range, using pairs of correlated visible and infrared photons *(1, 2)*. One of the most prominent demonstrations *(3)* was performed with 532 nm pump light generating photons at 1.5 µm and at 810 nm and is based on the effect of induced coherence without induced emission *(4, 5)*. The general principle of quantum sensing in the terahertz frequency range has previously been demonstrated by Kitaeva et al. *(6)*. They used a single-crystal interferometer in Young's configuration to measure the absorption of a periodically poled lithium niobate crystal (PPLN) in the terahertz frequency range. A different setup with a Mach-Zehnder geometry was used to measure the linear properties of a 4% magnesium oxide (MgO) doped lithium niobate ($LiNbO_3$) crystal *(7)*. To that end, two 5% MgO-doped $LiNbO_3$ crystals were placed right around the investigated object. However, all reported setups have in common, that the idler radiation is not coupled out of $LiNbO_3$, which is a crucial requirement for measurements of any external sample.

In our experiment, we generate terahertz (idler) photons via spontaneous parametric down-conversion (SPDC) using pump photons at 660 nm, which results in the generation of signal photons at a wavelength of about 661 nm, spectrally very close to the pump wavelength. Further, down- as well as up-conversion of thermal radiation plays a major role in the terahertz frequency range, as our previous work *(8)* has shown. Hence, to evaluate the feasibility of quantum sensing at room temperature, we first theoretically analyze our concept for a single-crystal quantum interferometer.

## Theoretical analysis

A schematic representation of the setup is shown in Fig. 1. A pump beam (not shown in Fig. 1) coherently illuminates a nonlinear crystal and creates pairs of signal (s) and idler (i) photons. Our theoretical description of this process is based on *(9, 10)*. In the single-mode approximation, the annihilation operators for the photons in the signal ($\hat{a}'_{s1}$) and idler ($\hat{a}'_{i1}$) output modes after the first passage through the crystal are given by

$$\hat{a}'_{s1} = u_1 \hat{a}_{s1} + v_1 \hat{a}^\dagger_{i1},$$
$$\hat{a}'_{i1} = u_1 \hat{a}_{i1} + v_1 \hat{a}^\dagger_{s1},$$

in terms of the input modes $\hat{a}_{s1}$ and $\hat{a}_{i1}$, respectively. The elements of the scattering matrix $u_1$ and $v_1$ correspond to the conversion rates, where $u_1$ describes up-conversion and $v_1$ together with $u_1$ down-conversion (9). In usual SPDC experiments, the input modes are in their vacuum states. However, in our case of idler photons in the terahertz range, due to the small energy, the idler inputs receive significant contributions from thermal fluctuations and are described to be in a thermal state. After the crystal, the pump and signal photons are separated from the idler photons, which interact with the object. The pump, signal, and idler radiation is then reflected and coupled back into the crystal. The signal ($\hat{a}'_{s2}$) and idler ($\hat{a}'_{i2}$) output modes after the second passage through the crystal read

$$\hat{a}'_{s2} = u_2 \hat{a}_{s2} + v_2 \hat{a}^\dagger_{i2},$$
$$\hat{a}'_{i2} = u_2 \hat{a}_{i2} + v_2 \hat{a}^\dagger_{s2},$$

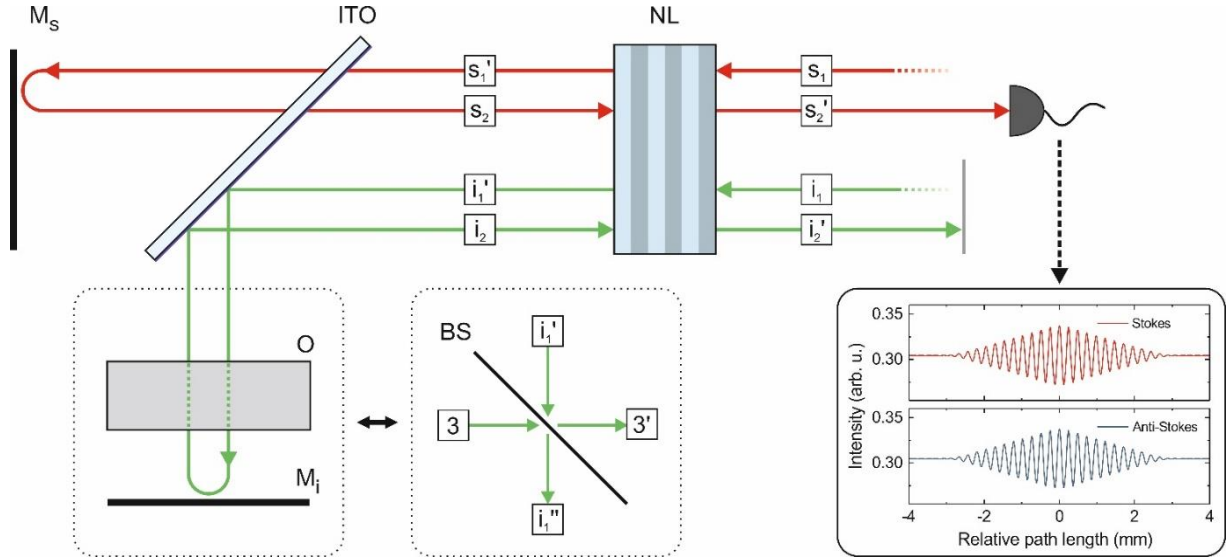

**Fig. 1. Scheme and nomenclature for the theoretical analysis.** In addition to a laser pump (for simplification not drawn here), the signal ($s_1$) and idler ($i_1$) input modes enter the nonlinear crystal (NL). The interaction in the crystal leads to the generation of signal and idler photons in the output modes $s'_1$ and $i'_1$, respectively. They are separated by an indium tin oxide coated (ITO) glass. Afterwards, the signal radiation and the pump beam are reflected back into the crystal by the mirror $M_s$. The input modes for the second passage are denoted by $i_2$ and $s_2$, which is, due to the alignment, equal to $s'_1$. The idler mode $i'_1$ passes through the object (O), is reflected by the mirror $M_i$, and propagates through the object again. This acts as a beam splitter (BS) with second input mode 3 and output modes $i''_1$ and $3'$. Aligning the idler beams, the mode $i''_1$ corresponds to $i_2$. The output modes after the second passage are $s'_2$ and $i'_2$. Finally, the signal radiation (in mode $s'_2$) is detected by the detector. The inset shows the simulated interference signal in the Stokes (red) and Anti-Stokes (blue) regions based on the detailed model (see Methods).

where $\hat{a}_{s2}$ and $\hat{a}_{i2}$ denote the signal and idler input modes for the second passage, respectively, and $u_2$ and $v_2$ are the conversion rates, as described above for $u_1$ and $v_1$. These transformations correspond to Bogoliubov transformations. When losses can be neglected, the transformations are unitary and thus obey

$$1 = |u_j|^2 - |v_j|^2 = U_j - V_j, \quad j = 1, 2.$$

The object placed in the idler path is modelled as a lossless beam splitter, described by the transformation

$$\hat{a}''_{i1} = t\hat{a}'_{i1} + r\hat{a}_3,$$

with complex transmissivity $t$ and reflectivity $r$, and $1 = |t|^2 + |r|^2 = T + R$. When the idler and signal beams are aligned in the crystal and thus become indistinguishable, the relation between the input and output modes is given by

$$\hat{a}_{i2} = \exp(i\phi_i)\hat{a}''_{i1},$$
$$\hat{a}_{s2} = \exp(i\phi_s)\hat{a}'_{s1},$$

where $\phi_i$ and $\phi_s$ denote the phases gained by the propagation of the modes between the crystal.

The observed signal count rate can then be calculated from

$$\hat{N}'_{s2} = \hat{a}'^\dagger_{s2}\hat{a}'_{s2} = |t^*v_1^*v_2 + u_1u_2|^2 \hat{N}_{s1} + RV_2(\hat{N}_3 + 1) + |t^*u_1^*v_2 + v_1u_2|^2 (\hat{N}_i + 1) + \text{mixed terms}.$$

Here, the phases $\phi_{i/s}$ have been absorbed in $u_2$ and $v_2$. While the signal input modes are in the vacuum state $\langle\hat{N}_{s1}\rangle = \langle\hat{N}_{s2}\rangle = 0$ and $\langle\hat{a}_j^\dagger\hat{a}_k\rangle = \langle\hat{a}_k\hat{a}_j^\dagger\rangle = 0$ for all $j \neq k$, the idler inputs are in thermal states $\langle\hat{N}_i\rangle = \langle\hat{N}_3\rangle = N_{th} = 1/[\exp(\hbar\omega_i/k_BT) - 1]$, where $\omega_i$ denotes the idler angular frequency and $T$ the temperature. In the case of equal gain in both passages, $V_1 = V_2 = V_0$, we obtain for the down-conversion signal count rate after the second traverse

$$R_s = \langle\hat{N}'_{s2}\rangle = (N_{th} + 1)2V_0\left[1 + \frac{1}{2}V_0T + \frac{1}{2}V_0 + \sqrt{T(1 + 2V_0 + V_0^2)}\cos(2\phi)\right],$$

where all phase terms are summarized in $\phi$.

Analogously, a similar expression can be derived for the case of up-conversion. Here, the signal is generated only due to thermal idler photons and the vacuum fluctuations do not contribute. Our experimental setup operates in the low-gain regime ($V_0 \ll 1$) and contains a nearly transparent object (for simplification $T = 1$) of thickness $d$ with refractive index $n$. Furthermore, the signal and pump paths are fixed, while the mirror in the idler path is translated by the distance $x$. In this case, the observed signal rate is given by

$$R_s = (N_{th} + 1)2V_0\left[1 + \cos\left(\phi_0 + \frac{\omega_i}{c}[2x + (n-1)2d]\right)\right],$$

where constant phase terms are summarized in $\phi_0$. While the derivation presented so far has been performed for the single-mode case, we also developed a theoretical multi-mode model, in order to better incorporate the experimental circumstances, see Methods for further details. The expected interference resulting from the model is displayed in Fig. 1. Hence, we conclude that the observation of an interference pattern can be expected in the presence of thermal photons for down- and even for up-conversion.

**Experimental setup**

Our experimental setup shown in Fig. 2 bases on a previously presented setup *(8)*, which is extended to a Michelson-like single-crystal quantum interferometer *(11, 12, 13)*. A 660 nm frequency-doubled solid-state laser is used as pump source and the pump photons are coupled to the interferometer part using a volume Bragg grating. As nonlinear medium, a 1 mm long periodically poled $LiNbO_3$ crystal with a poling period of 90 µm is used, where visible (signal) photons and associated (idler) photons in the terahertz frequency range are generated. Behind the crystal, an indium tin oxide (ITO) coated glass is placed, separating the idler photons from the pump and signal photons. The pump and signal radiation is directly focused back into the crystal by a concave mirror. The high refractive index of $LiNbO_3$ in the terahertz frequency range leads to a large scattering angle of the idler radiation. Therefore, the idler radiation is first collimated by a parabolic mirror and afterwards reflected at a plane mirror, placed on a piezoelectric linear stage. After two passages through the crystal, the pump and signal beams are collimated and the pump photons are filtered by three volume Bragg gratings acting as highly efficient and narrowband notch filters. An uncooled scientific complementary metal-oxide-semiconductor (sCMOS) camera is used as detector.

The inset in Fig. 2 shows the observed frequency-angular spectrum of the used crystal. The transmission grating leads to a distribution of different signal wavelengths (corresponding to the generated terahertz frequency) in the x-direction of the camera, while the angular distribution is observed along the y-direction. We observe the tails in the Stokes (down-conversion) as well as in the Anti-Stokes (up-conversion) region, which are typical for $LiNbO_3$ in the terahertz frequency range. The signal photons are generated either by SPDC or by conversion of thermal photons in the terahertz frequency range *(14)*. The highest count rates are observed for the collinear parts of all tails, originating from the first-order forward and backward generation in the Stokes and Anti-Stokes cases. Calculating the frequency shift, which corresponds to ideal phase matching for the poling period of the crystal, we obtain 1.26 THz in collinear forward and 0.47 THz in collinear backward direction, matching the observed spectra. The image is a superposition of the frequency-angular spectra generated on the way back and forth through the crystal. In addition, the signal intensity is linearly dependent on the pump power (see Methods, Fig. 5) meaning the experiment performs in the low-gain regime.

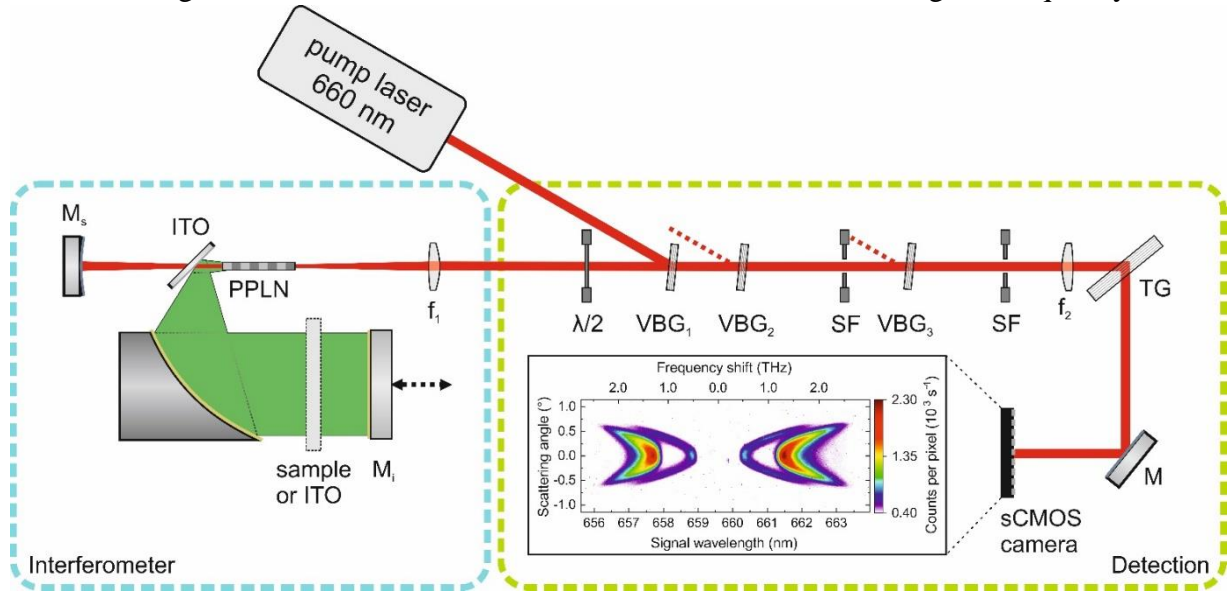

**Fig. 2. Schematic of the experimental setup.** A continuous wave laser with a wavelength of 659.58 nm is reflected by a volume Bragg grating ($VBG_1$) into the interferometer part of the setup through a zero-order half-wave plate ($\lambda/2$) controlling the polarization. It is then focused by a lens $f_1$ into a periodically poled 1 mm long MgO-doped $LiNbO_3$ (PPLN) crystal generating signal and terahertz photons that are separated by an ITO. Signal and pump radiation are reflected at $M_s$ directly into the crystal. The terahertz radiation passes the object twice being reflected by a moveable mirror $M_i$. In the second traverse of the pump through the PPLN, additional signal and idler photons are generated. Afterwards, the lens $f_1$ collimates the pump and signal radiation for the detection starting with filtering the pump radiation by three VBGs and spatial filters (SF). To obtain the frequency-angular spectrum, the signal radiation is focused through a transmission grating (TG) by the lens $f_2$ onto a sCMOS camera. The inset shows a frequency-angular spectrum for the used crystal (poling period $\Lambda = 90$ µm, pumped with 450 mW). The scattering angle corresponds to the angle after the transmission from the crystal to air.

**Results**

The dependence of the signal photon count rate in collinear forward direction on the position of the translation stage is shown in Fig. 3 (A and B top). The data is acquired by integrating the collinear forward regions of the frequency-angular spectra for different lengths of the terahertz (idler) path (see Methods for more details). Interference of the signal photons is observed in the Stokes as well as the Anti-Stokes region, matching the simulated interference signal (see Fig. 1). The corresponding FFTs show a peak in both cases at a frequency of about 1.26 THz (see Fig. 3C and 3D) in accordance to the phase-matching conditions. The noise of the recorded data is mainly due to laser fluctuations and noise of the camera. To ensure that the interference is caused by terahertz photons propagating along the idler path, an additional stationary ITO glass is placed between the parabolic and the plane mirror. The terahertz radiation is blocked at the ITO glass, while visible light is mainly transmitted (bottom of Fig. 3A and 3B). In this case, no interference is observed and the FFTs show no significant peak.

To demonstrate terahertz quantum sensing, we measure the thickness of various polytetrafluoroethylene (PTFE) plates with a maximum thickness of 5 mm placed in the idler path. Due to the refractive index of PTFE, the optical length of the path changes and the envelope of the interference is observed at different stage positions (see Fig. 4A and 4B). In addition to the shift, the visibility of the interference decreases when the PTFE plate is added due to Fresnel reflections at the surfaces. To determine the thickness of the plates, we estimated their refractive index using a standard time-domain spectroscopy (TDS) system *(15)* to be $1.42 \pm 0.01$. Knowing the refractive index and the shift of the interference signal allows for the calculation of the layer thickness. To specify this shift, we performed a fit (see Methods) of the signal with and without plate and determined the central position of the interference. In Fig. 4C, the thicknesses measured by quantum interference are plotted over the thicknesses measured with a micrometer caliper. The experimental results show clearly that quantum interference with idler photons in the terahertz frequency range enables the determination of layer thicknesses of samples placed in the terahertz path.

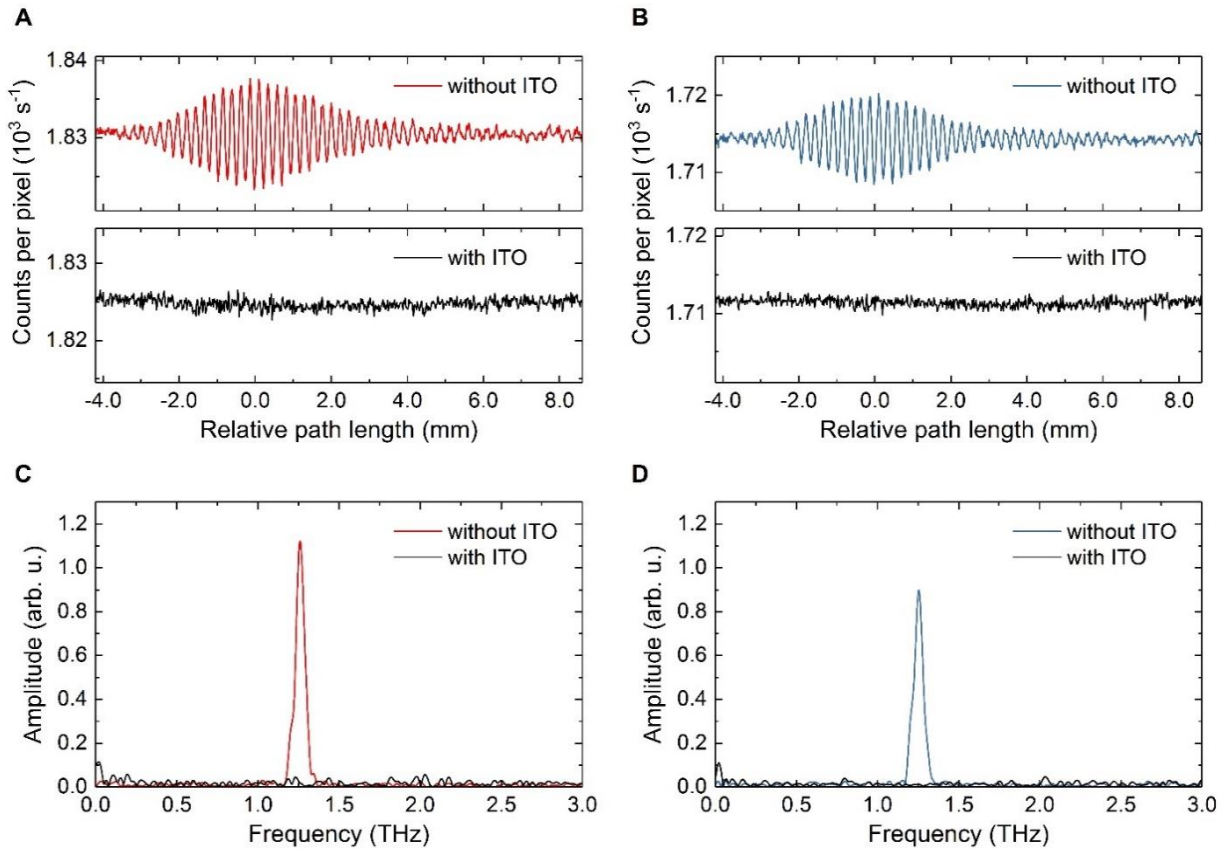

**Fig. 3. Terahertz quantum interference.** In the collinear forward spot of the signal, interference is observed in the **A** Stokes as well as in the **B** Anti-Stokes region. **C** and **D** show the corresponding FFTs peaks at about 1.26 THz. By placing an additional ITO glass in the idler path, no interference can be observed and the peaks in the FFTs disappear.

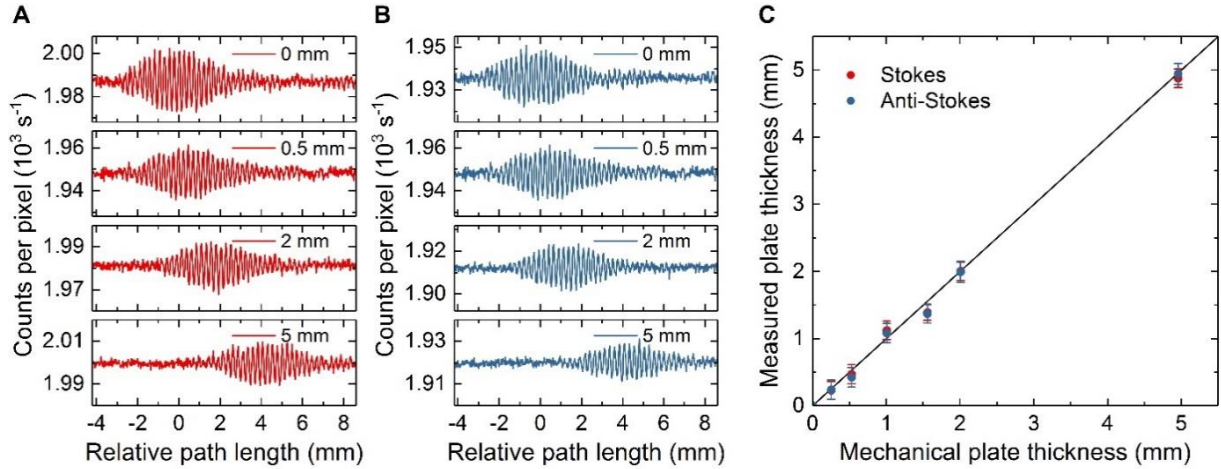

**Fig. 4. Terahertz quantum sensing.** The envelope of the interference is shifted depending on the thickness of the PTFE plate in the **A** Stokes- and **B** Anti-Stokes part. **C** Thickness of the PTFE plate measured by quantum interference over PTFE thickness measured by a micrometer caliper. The solid line is the angle bisector. The horizontal error bars (hidden by the data points) take into account the uneven thicknesses of the PTFE plates as well as the inaccuracy of the reference measurement. The vertical error bars result from the precision of determining the shift of the envelope center of the interference.

**Discussion and conclusion**

A common explanation for the observed interference is based on the principle of indistingu-ishability of the generated photon pairs *(3, 5, 16)*. They are created either in the first or second traverse of the pump radiation through the nonlinear crystal. If the paths of the idler and the signal photons are perfectly aligned, the information on the path of origin is lost. Therefore, the paths are indistinguishable and the photons interfere. On the other hand, also classical interpretations exist to explain the effect *(17-19)*. The quantum-mechanical interpretations *(3, 5, 16)* were tailored to SPDC processes. In our experiment, operating in the strongly frequency non-degenerated regime, further thermal radiation exists, creating additional signal photons due to sum- (up-conversion) and difference-frequency generation (down-conversion). However, as thermal photons have no phase correlation, a classical explanation of the observed interference is not possible. On the other hand, the principle of indistinguishability applies also to this case and thus explains the observed interference, therefore, being regarded as a quantum-mechanical explanation.

In conclusion, we observed quantum interference in the terahertz frequency range with propagation of terahertz photons in free space. This interference is observable in the Stokes as well as in the Anti-Stokes region. Additionally, we demonstrated the capability to use this technique to determine the thicknesses of various PTFE plates, as layer thickness measurements are one of the most prominent applications in the terahertz frequency range. Even though, the achieved measurement time and resolution are not yet comparable to classical terahertz measurement schemes, the presented demonstration of this concept is a first milestone towards terahertz quantum imaging.

## Materials and Methods

### Experimental Design

**Experimental interferometer** As pump source in the experiment, we use a linearly polarized, frequency-doubled, single-longitudinal-mode cw laser (Cobolt Flamenco) at a wavelength of 659.58 nm, providing a narrow bandwidth of less than 1 MHz at an average output power of up to 500 mW. A volume Bragg grating (BNF.660 from OptiGrate) reflects the laser into the interferometric part of the setup. For the generation of correlated photon pairs, a short 5×1×1 mm$^3$ LiNbO$_3$ crystal with a quasi phase matching (QPM) structure (PPLN) is used. The crystal is structured in the 1-mm direction, which is also the direction of the pump beam. For the creation of signal and idler photons, energy and momentum conservation has to be fulfilled. The corresponding phase-matching considerations of LiNbO$_3$ are shown in detail in a previous work *(8)*.

In our experiment, we use a single-crystal Michelson-like setup (see Fig. 2). The advantage of this setup is that just one crystal is used. Usually, two different crystals are used. However, the two crystals may have minor deviations in their parameters, which have a negative influence on the visibility of the interference. Due to the high absorption in the terahertz frequency range, longer PPLN crystals are disadvantageous for terahertz applications *(20)*.

In order to align the polarization of the pump radiation, a zero-order half-wave plate at the design wavelength of 670 nm is placed in the beam path, ensuring extraordinary pump polarization and type 0 phase matching in the PPLN crystal.

A decent focusing into the crystal is achieved using a lens with a focal length of 200 mm placed 200 mm in front of the crystal. After the crystal, pump and signal photons are separated from the idler photons with an ITO glass placed 7 mm behind the crystal at an angle of 45 degrees (to the propagation axis of the pump beam). Pump and signal photons are focused back into the nonlinear crystal by a concave mirror with a focal length of 100 mm placed at a distance of 200 mm to the crystal. Because of the high refractive index of LiNbO$_3$ in the terahertz region, the idler radiation is emitted at large scattering angles. To collect a major part of the idler photons a 2-inch parabolic mirror with a focal length of 50.8 mm is used to collimate the idler photons. After collimation, the idler radiation is reflected by a gold-coated plane mirror and focused back into the nonlinear crystal again by the parabolic mirror. The plane gold-coated mirror is placed on a coarse manual linear stage (resolution ± 5 µm) and an automated fine translation stage (resolution ± 8 nm). The latter is the linear stage Q-522.030 from PI with a displacement up to 6.4 mm and a minimum increment of 8 nm.

**Detection** In order to detect the created signal photons, the pump and signal radiation is collimated by the former focusing lens, now acting as a collimating lens. After collimation, the pump radiation is filtered by three volume Bragg gratings (separated by a distance of 1 m from one another) acting as highly efficient notch filters. They efficiently reflect the pump photons while transmitting the down- and up-converted signal photons having a frequency that is slightly shifted. All volume Bragg gratings exceed an optical density of 4 at the design wavelength and transmit more than 90% already at a frequency shift of 0.1 THz *(8)*. Spatial filters block divergent stray light originating from the VBGs.

To spectrally resolve the generated signal photons, a highly efficient transmission grating (PCG-1908-675-972 from Ibsen photonics with 1908 lines per millimeter, efficiency greater than 94% at the used wavelength) is employed. A lens with a focal length of 400 mm is placed 35 mm in front of the transmission grating. The distance between the transmission grating and the detector camera is 365 mm, so the distance between the focusing lens and the camera matches the focal length of the lens. For signal photon detection, an uncooled scientific CMOS camera (Thorlabs Quantalux™ sCMOS Camera) with a specified quantum efficiency of about 55% (at 660 nm) is used. Its pixel size is 5.04×5.04 µm², providing 2.1 megapixels with up to 87 dB dynamic range. At 20 °C, the labeled pixel dark count rate and readout noise is about 20 counts per second and 1 e⁻, respectively. The measured background illumination of the presented results is about 160 counts per second and pixel (combining dark count rate, remaining stray and ambient light).

**Frequency-angular spectrum** The combination of the lens $f_2$ and the transmission grating leads to a frequency-angular spectrum on the camera (see Fig. 2). The data for the frequency-angular spectrum were acquired with an illumination time of 4 s and a pump power of 450 mW. At room temperature, one can observe the first-order forward and backward generation in the Stokes as well as in the Anti-Stokes region. Due to SPDC, the Stokes region receives a higher count rate than the Anti-Stokes region, where only conversion of thermal photons contributes. Due to apertures of the wave plate and filters along the beam path, the spectrum shows a vertical limitation starting from a scattering angle at about 0.5 degrees. A more detailed description is given in a previous work *(8)*.

**Quantum sensing** To perform the quantum-sensing measurement, we changed the path length of the terahertz idler radiation by moving the position of the fine translation stage with a step size of 10 µm over the whole stage width (6.4 mm). This leads to a change of the idler optical path length of 20 µm at every step. At each position, an image was recorded with an exposure time of 500 ms leading to a total measurement time of about 10 minutes. Subsequently, we take the average over an area of 25×10 pixels in the collinear regions of the first-order forward tails in the Stokes and Anti-Stokes regions for every image. To minimize the influence of ambient light and dark count rate, we additionally averaged the count rate of a same size area without signal, and subtracted it from the measured value. Because a single measurement has a low signal-to-noise ratio, the procedure is repeated 30 times and averaged again. In Fig. 4, the mean Stokes and Anti-Stokes count rates of the measurements with the various PTFE plates show small variations. These are caused by little temperature changes, as on the one hand, the number of thermal photons is highly temperature dependent and, therefore, the conversion rate increases with the temperature. On the other hand, the phase-matching conditions slightly changes, moving the position of the collinear forward region.

**Analysis of the quantum-sensing measurement values** To determine the thickness of the PTFE plates, first, we performed a measurement without an object in the idler path as reference. Inserting PTFE plates changes the effective refractive index in the idler path and the interference signal is shifted. To calculate the plate thickness, in addition to the shift, the refractive index of the plate must be known.

The index of refraction is estimated to be $n_{\text{PTFE}} = 1.42 \pm 0.01$ with a standard TDS-system, measuring the time between the reflection on the front and backside of the PTFE plate. To access the quality of the quantum-sensing measurements, the thickness of the used plates was measured with a micrometer caliper at 10 different positions and averaged.

To determine the shift due to the PTFE plate, we fitted the following function to the measured interference pattern:

$$f(x) = y_0 + A \cdot \sin(\nu \cdot x + \varphi) \cdot \exp\left(-\frac{(x-x_c)^2}{2\omega^2}\right).$$

In the used function, $y_0$ is the average count rate and $A$ is the amplitude of the envelope. As a result, $A/y_0$ is the visibility of the signal, $\nu$ is the angular frequency of the interference and $\varphi$ is the phase of the sinusoidal signal. Finally, $x_c$ is the center of the envelope and $\omega$ represents its width. This function takes into account the sinusoidal interference pattern and the envelope, but neglecting the asymmetry of the signal. However, this simplification is acceptable for determining the thickness of the PTFE plates especially as the asymmetry is very similar for each measurement.

Finally, from the index of refraction of the PTFE plates and the difference between the centers of the envelopes of the reference measurement and the measurement with PTFE plates in the idler beam path, the plate thickness can be calculated.

The horizontal error bars provided in Fig. 4C result from the inaccuracy of the TDS-measurements as well as the uneven thicknesses of the PTFE plates over the tested areas. However, the error bars are so small that they are hidden by the data points. On the other hand, the vertical error bars take into account the errors that arise when fitting $x_c$ to the measured data. As an example, for the 5.0 mm PTFE plate the relative accuracy is 2.9 % for the Stokes and 3.1 % for the Anti-Stokes measurement.

**Excluding induced emission** To prove that no stimulated emission takes place during the second traverse of the radiation generated in the first pass through the crystal, the signal count rate is compared with and without the idler path blocked. To that end, 500 single measurements with a relative idler path length of 8.2 mm with respect to Fig. 3 are averaged for both cases and different pump powers. The corresponding data are plotted in Fig. 5 and show nearly the same count rates for both cases, in the Stokes as well as in the Anti-Stokes region. Hence, the ratio is about 1.0. This is a clear indication, that photons generated in the first passage do not cause noticeable induced emission. Furthermore, Fig. 5 shows a linear dependency of the signal count rate on the pump power, indicating that the experiment is performed in the low-gain regime.

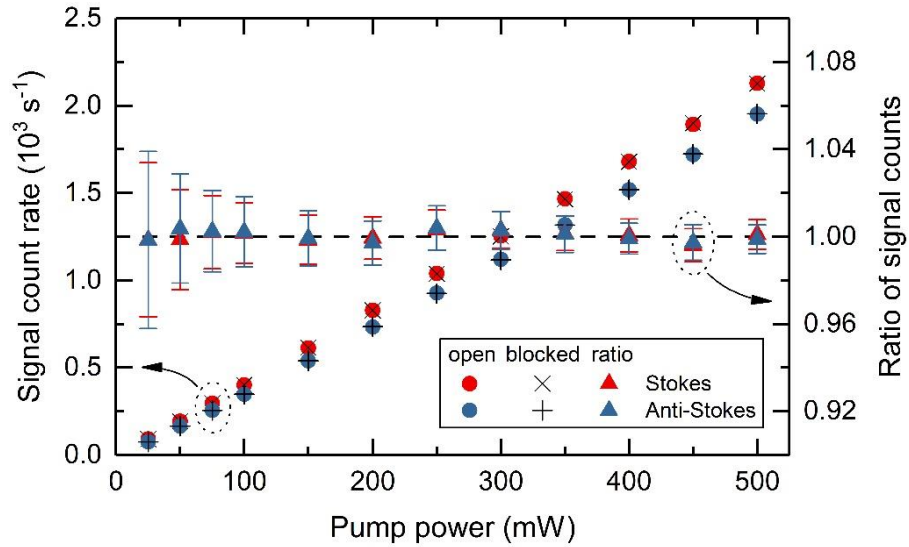

**Fig. 5. Investigation of induced emission by idler radiation generated in the first passage.** The count rates of the collinear Stokes (red, cross) and Anti-Stokes (blue, plus) region are shown for the cases of the idler beam being blocked (crosses) and unblocked (colored dots) for various pump powers. The triangles are the ratios between the unblocked and blocked cases. All values are close to a ratio of 1.00 (dashed line), which lies in the range of all error bars.

## Theoretical analysis

**Theoretical model** In the following, we present a theoretical calculation of the interference signal. While the derivation in the main part of the paper has been performed for the single-mode case, we now provide a multi-mode description, which better fits the experimental situation.

To that end, we follow the approach in *(5, 21, 22)*. The quantum state after the second passage through the crystal can be described as a coherent superposition of the two-photon states produced in the first and the second passage

$$|\Psi\rangle = N_1 \int d^3k_{s1} d^3k_{i1} \Phi(\mathbf{k}_{s1}, \mathbf{k}_{i1}) \hat{a}^\dagger(\mathbf{k}_{s1}) \hat{a}^\dagger(\mathbf{k}_{i1}) |0\rangle + N_2 \int d^3k_{s2} d^3k_{i2} \Phi(\mathbf{k}_{s2}, \mathbf{k}_{i2}) \hat{a}^\dagger(\mathbf{k}_{s2}) \hat{a}^\dagger(\mathbf{k}_{i2}) |0\rangle,$$

where $\mathbf{k}_{sj}$ and $\mathbf{k}_{ij}$, $j=1,2$, denote the signal and idler wave vectors in $j$th passage through the crystal, respectively, and $N_1$, $N_2$ are constants describing the gain in the $j$th passage. In our case of equal gain $|N_1|=|N_2|$.

The two-photon states are calculated from first-order perturbation theory for an undepleted monochromatic collimated Gaussian beam propagating in $z$ direction. The employed Hamiltonian and further details can be found in literature *(8)*. We take into account the finite extent of the crystal yielding

$$\Phi(\mathbf{k}_s,\mathbf{k}_i) = \delta(\omega_p - \omega_s - \omega_i) \exp\left(-\frac{w^2}{4}(\mathbf{k}_{s\perp}+\mathbf{k}_{i\perp})^2\right) \mathrm{sinc}\left(\frac{\Delta k_z L}{2}\right),$$

where $\omega_{p,s,i}$ denote the angular frequency of pump (p), signal (s), and idler (i), $\mathbf{k}_{s,i\perp}$ are the transverse components of the wave vectors, $w$ is the pump beam waist, $L$ the crystal length, and $\Delta k_z = k_{pz} - k_{sz} - k_{iz} + \frac{2\pi}{\Lambda}$, with the poling period $\Lambda$.

The positive-frequency part of the electric field at the detector is given by

$$\hat{E}_s^{(+)}(\mathbf{r}) = \hat{a}_{s1}(\mathbf{k}_s) + \exp(i\phi_s)\hat{a}_{s2}(\mathbf{k}_s),$$

and the signal count rate is proportional to

$$R(\mathbf{r}) = \langle\Psi|\hat{E}_s^{(-)}(\mathbf{r})\hat{E}_s^{(+)}(\mathbf{r})|\Psi\rangle,$$

where $\hat{E}_s^{(-)}(\mathbf{r}) = \left(\hat{E}_s^{(+)}(\mathbf{r})\right)^\dagger$. The interaction of the idler with the object and the subsequent alignment of the idler modes in the crystal can be modeled as a beam splitter and expressed as

$$\hat{a}_{i2}(\mathbf{k}_i) = \left[t(\mathbf{k}_i)\hat{a}_{i1}(\mathbf{k}_i) + r'(\mathbf{k}_i)\hat{a}_{i0}(\mathbf{k}_i)\right]\exp[i\phi_i(\mathbf{k}_i)],$$

for the case of perfect alignment of the idler beams. Here, $t(\mathbf{k}_i)$ is the transmission coefficient of the object, $r'(\mathbf{k}_i)$ the reflection coefficient when illuminated from the opposite direction, $\hat{a}_{i0}(\mathbf{k}_i)$ the vacuum input at the unused port of the beam splitter, and $\phi_i(\mathbf{k}_i)$ is the phase gained by the mode $\mathbf{k}_i$. We work in the diffraction-less limit, where there is a one-to-one correspondence between a point on the object or detector and the associated idler or signal wave vector, respectively. Furthermore, we have neglected the absorption within the crystal and also the propagation of the idler and signal modes inside the crystal.

After some algebra, we obtain for the signal rate in the detector plane ($xy$-plane)

$$R(x,y) = N\int d\omega_s d^3k_i \delta(\omega_p - \omega_s - \omega_i)\exp\left(-\frac{w^2}{2}(\mathbf{k}_{s\perp}+\mathbf{k}_{i\perp})^2\right)\mathrm{sinc}^2\left(\frac{\Delta k_z L}{2}\right) \quad (1)$$
$$\times\left[1 + t(\mathbf{k}_{i\perp})\cos(\phi_i(\mathbf{k}_i))\right],$$

where constants and slowly-varying parameters, e.g., angular frequencies $\omega_{p,s,i}$ or refractive indices $n_{p,s,i}$, are summarized in $N$.

**Calculation of the interference signal** We express the integral in Eq. 1 in spherical coordinates and employ the paraxial approximation, since our following calculation focusses on the collinear signal region, where also the corresponding idler angles are not too large. Since, in the current experiment, the object we want to describe has no transversal structure, we only keep a $\theta_i$ dependent transmission and phase term. In this way, later, we are able to include limiting effects of the imaging system. Then, the count rate reads

$$R(\theta_s, \varphi_s) = N \int d\omega_i \theta_i d\theta_i d\varphi_i \exp\left(-\frac{w^2}{2}\left[(k_s\theta_s)^2 + (k_i\theta_i)^2 + 2k_s\theta_s k_i\theta_i \cos(\varphi_s - \varphi_i)\right]\right)$$

$$\times \operatorname{sinc}^2\left(\frac{\Delta k_z L}{2}\right)[1 + t(\theta_i)\cos(\phi_i)],$$

with $\Delta k_z = k_p - k_s(1 - \frac{1}{2}\theta_s^2) - k_i(1 - \frac{1}{2}\theta_i^2) + \frac{2\pi}{\Lambda}$, $k_j = n_e(\omega_j)\frac{\omega_j}{c}$, for $j$=p,s,i, and $n_e(\omega_j)$ are the extraordinary refractive indices. Here and in the following, constant parameters are summarized in the constants $N'$ ($N''$,...). We have also carried out the integration over $\omega_s$ such that, from now on, $\omega_s = \omega_p - \omega_i$ has to be used in the formulae.

Neglecting the angular dependence of the refractive indices of PPLN, which is valid in the paraxial regime, we carry out the integration over $\varphi_i$ and obtain

$$R(\theta_s) = N \int d\omega_i \theta_i d\theta_i \exp\left(-\frac{w^2}{2}\left[(k_s\theta_s)^2 + (k_i\theta_i)^2\right]\right) I_0(w^2 k_s\theta_s k_i\theta_i) \operatorname{sinc}^2\left(\frac{\Delta k_z L}{2}\right)(1 + t(\theta_i)\cos(\phi_i)),$$

with the modified Bessel function of the first kind $I_0(x)$. The large difference of the signal and idler frequencies results in idler scattering angles, which are much larger than the corresponding signal angles. This can be seen in Fig. 6, where the idler angular density corresponding to collinear signal angles, $\theta_s = 0$, is plotted for three different pump beam waists. Here, the angular density has been evaluated using the refractive indices *(23, 24)* for 5 mol.% MgO-doped LiNbO$_3$ and the crystal parameters are $L$=1 mm and $\Lambda = 90$ µm.

Since the pump width is finite, the transversal momentum conservation is not perfect and $\Gamma(\theta_i)$ receives contributions also for idler angles $\theta_i > 0$.

Moreover, due to the large refractive index of PPLN in the terahertz range and the resulting total internal reflection, only idler angles up to 11 degrees inside the crystal can be coupled out and aligned in the second passage through the crystal. This maximal idler angle is further limited by the apertures of the optical elements in the idler path. As a result, we expect to observe interference only in the collinear part of the signal spectrum and a reduction of the visibility, since not all associated idler modes can be aligned and made indistinguishable. To calculate the interference signal under variation of the idler path length $\Delta l_i$, we evaluate $R(\theta_s)$ for collinear signal angles, i.e. $\theta_s = 0$, and model the idler angles, which can be coupled out by a transmission that takes the form of the step function $t(\theta_i) = \Theta(\theta_i^{max} - \theta_i)$:

$$R(0) = N \int d\omega_i \theta_i d\theta_i \exp\left(-\frac{w^2}{2}(k_i\theta_i)^2\right) \text{sinc}^2\left(\frac{\Delta k_z L}{2}\right)\left[1 + \Theta(\theta_i^{max} - \theta_i)\cos\left(\phi_0 + \frac{\omega_i}{c}\Delta l_i\right)\right], \quad (2)$$

where $\phi_0$ summarizes constant phase terms. An analogous expression can holds for the case of up-conversion. To that end, in Eq. 2, we only need to change the energy-conservation condition to $\omega_s = \omega_p + \omega_i$ and the longitudinal phase mismatch takes the form $\Delta k_z = k_p - k_s(1-\tfrac{1}{2}\theta_s^2) + k_i(1-\tfrac{1}{2}\theta_i^2) - \tfrac{2\pi}{\Lambda}$. The results for $R(0)$, in the cases of down- and up-conversion, are plotted in Fig. 1 as a function of the relative idler path length. For the calculation, we employed the crystal parameters and refractive indices given above and $w$=60 µm. The maximal idler angle which can be aligned in the second passage is given by the aperture of the parabolic mirror and corresponds to $\theta_i^{max}$ =5° inside the crystal.

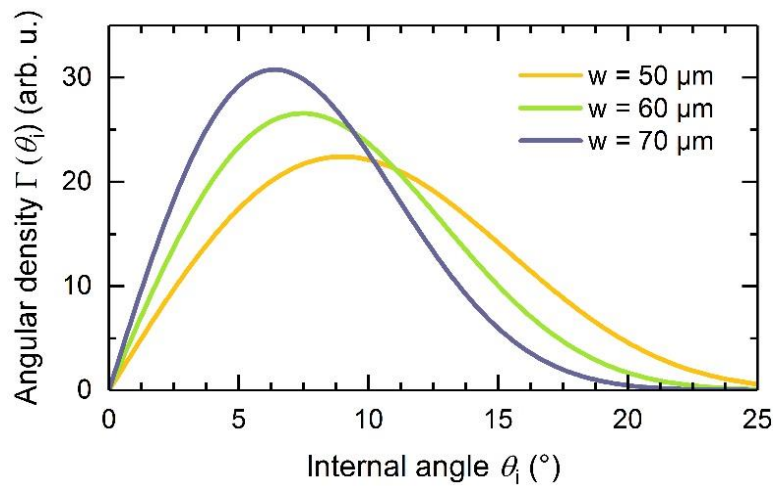

**Fig. 6. Idler angular distribution:** The idler angular density $\Gamma(\theta_i)$ corresponding to collinear signal angles is plotted as a function of the idler angle inside the PPLN crystal $\theta_i$ for three different pump radii.

The comparison with the experimental results in Fig. 3 shows that the shapes and coherence lengths are similar (within the fluctuations). Due to the fact, that not all idler photons can be aligned in the crystal, the visibility of the theoretical signal is already only about 30%. Further effects substantially limiting the visibility, which are not included in the simulation, are the absorption of terahertz photons in PPLN, Fresnel reflection losses, the propagation of the photons inside the crystal. In addition, in contrast to the simulated signal, the measured interference is slightly asymmetric, which is subject of current investigation.

**References and Notes**

1. M. V. Chekhova, Z. Y. Ou, Nonlinear interferometers in quantum optics. *Adv. Opt. Photon.* **8**, 104-155 (2016).
2. D. A. Kalashnikov, A. V. Paterova, S. P. Kulik, L. A. Krivitsky, Infrared spectroscopy with visible light. *Nat. Photonics* **10**, 98-101 (2016).
3. G. B. Lemos, V. Borish, G. D. Cole, S. Ramelow, R. Lapkiewicz, A. Zeilinger, Quantum imaging with undetected photons. *Nature* **512**, 409-412 (2014).
4. L. J. Wang, X. Y. Zou, L. Mandel, Induced coherence without induced emission. *Phys. Rev. A* **44**, 4614 (1991).
5. X. Y. Zou, L. J. Wang, L. Mandel, Induced coherence and indistinguishability in optical interference. *Phys. Rev. Lett.* **67**, 318-321 (1991).
6. G. K. Kitaeva, S. P. Kovalev, A. N. Penin, A. N. Tuchak, P. V. Yakunin, A method of calibration of terahertz wave brightness under nonlinear-optical detection. *J. Infrared Millim. Terahertz Waves* **32**, 1144-1156 (2011).
7. E. I. Malkova, S. P. Kovalev, K. A. Kuznetsov, G. K. Kitaeva, Nonlinear quantum interferometry in terahertz spectroscopy. *EPJ Web of Conferences* **195**, 06020 (2018).
8. B. Haase, M. Kutas, F. Riexinger, P. Bickert, A. Keil, D. Molter, M. Bortz, G. von Freymann, Spontaneous parametric down-conversion of photons at 660 nm to the terahertz and sub-terahertz frequency range. *Opt. Express* **27**, 7458-7468 (2019).
9. A. V. Belinsky, D. N. Klyshko, Interference of classical and non-classical light. *Phys. Lett. A* **166**, 303-307 (1992).
10. H. M. Wiseman, K. Mølmer, Induced coherence with and without induced emission. *Phys. Lett. A* **270**, 245-248 (2000).
11. A. Paterova, H. Yang, C. An, D. Kalashnikov, L. Krivitsky, Measurement of infrared optical constants with visible photons. *New J. Phys.* **20**, 043015 (2018).
12. M. G. Basset, J. R. León Torres, M. Gräfe, Compact quantum imaging based on induced coherence. *CLEO/QELS conference* **2019**, FF3D.6 (2019).
13. C. Lindner *et al.*, https://arxiv.org/abs/1909.06864v2 (2019).
14. G. K. Kitaeva, P. V. Yakunin, V. V. Kornienko, A. N. Penin, Absolute brightness measurements in the terahertz frequency range using vacuum and thermal fluctuations as references. *Appl. Phys. B* **116**, 929-937 (2014).
15. D. Molter, M. Trierweiler, F. Ellrich, J. Jonuscheit, G. von Freymann, Interferometry-aided terahertz time-domain spectroscopy. *Opt. Express* **25**, 7547-7558 (2017).
16. M. Lahiri, R. Lapkiewicz, G. B. Lemos, A. Zeilinger, Theory of quantum imaging with undetected photons. *Phys. Rev. A* **92**, 013832 (2015).
17. J. H. Shapiro, D. Venkatraman, F. N. Wong, Classical imaging with undetected photons. *Sci. Rep.* **5**, 10329 (2015).
18. A. C. Cardoso, L. P. Berruezo, D. F. Ávila, G. B. Lemos, W. M. Pimenta, C. H. Monken, P. L. Saldanha, S. Pádua, Classical Imaging with Undetected Light. *Phys. Rev. A* **97**, 033827 (2018).
19. M. Lahiri, A. Hochrainer, R. Lapkiewicz, G. B. Lemos, A. Zeilinger, Nonclassicality of induced coherence without induced emission. *Phys. Rev. A* **100**, 053839 (2019).
20. M. Unferdorben, Z. Szaller, I. Hajdara, J. Hebling, L. Pálfalvi, Measurement of refractive index and absorption coefficient of congruent and stoichiometric lithium niobate in the terahertz range, *J. Infrared Millim. Terahz Waves* **36**, 1203-1209 (2015).
21. M. Lahiri, R. Lapkiewicz, G. B. Lemos, A. Zeilinger, Theory of quantum imaging with undetected photons. *Phys. Rev. A* **92**, 013832 (2015).
22. M. Lahiri, A. Hochrainer, R. Lapkiewicz, G. B. Lemos, A. Zeilinger, Twin-photon correlations in single-photon interference. *Phys. Rev. A* **96**, 013822 (2017).

23. O. Gayer, Z. Sacks, E. Galun, A. Arie, Erratum to Temperature and wavelength dependent refractive indexequations for MgO-doped congruent and stoichiometric LiNbO$_3$. *Appl. Phys. B* **101**, 343–348 (2010).
24. X. Wu, C. Zhou, W. R. Huang, F. Ahr, F. X. Kärtner, Temperature dependent refractive index and absorptioncoefficient of congruent lithium niobate crystals in the terahertz range. *Opt. Express* **23**, 29729–29737 (2015).

**Acknowledgments**

**Acknowledgments:** We thank J. Klier for determining the index of refraction in the terahertz range of the used PTFE plates and M. Bortz for helpful feedback during this project.

**Funding:** This project was funded by the Fraunhofer-Gesellschaft within the Fraunhofer Lighthouse Project Quantum Methods for Advanced Imaging Solutions (QUILT).

**Author contributions:** G.v.F. and D.M. initiated this research. M.K., B.H. and D.M. designed the experiment. M.K. and B.H. carried out the experiment. P.B. and F.R. performed the theoretical analysis. All authors discussed the results and contributed to the writing of the manuscript.

**Competing interests:** The authors declare that they have no competing interests.

**Data and materials availability:** All data needed to evaluate the conclusions in the paper are present in the paper. Additional data related to this paper may be requested from the authors.